\documentclass[
    final            % use final for the camera ready runs
    ,a4paper,numcites
  ]
  {aipproc}
\setcitestyle{numbers}
\layoutstyle{8x11double}

\usepackage{amsmath,amssymb,subfigure,wrapfig,verbatim}

\newcommand{\be}{\begin{equation}}
\newcommand{\ee}{\end{equation}}
\newcommand{\bea}{\begin{eqnarray}}
\newcommand{\eea}{\end{eqnarray}}
\newcommand{\hk}{\hspace{0.1cm}}

\newcommand{\rk}{\right)}
\newcommand{\lk}{\left(}

\def\cb{\bar{c}} 
\def\D{\mathcal{D}}

\def\Tr{\mbox{Tr}} 
 
\def\G{\Gamma}

\newcommand{\Det}{\mbox{Det}}

\begin{document}

\title{\begin{center}Hamiltonian Dyson-Schwinger and FRG Flow Equations \\ of Yang-Mills Theory in Coulomb Gauge \end{center}}


\author{Hugo Reinhardt}{
  address={Universit\"at T\"ubingen, Institut f\"ur Theoretische Physik, Auf der Morgenstelle 14, 72076 T\"ubingen, Germany}
}

\author{Markus Leder}{
  address={Universit\"at T\"ubingen, Institut f\"ur Theoretische Physik, Auf der Morgenstelle 14, 72076 T\"ubingen, Germany}
}

\author{Jan M.~Pawlowski}{
  address={Universit\"at Heidelberg, Institut f\"ur Theoretische Physik, Philosophenweg 16, D-69120 Heidelberg, Germany\\
ExtreMe Matter Institute EMMI, GSI, Planckstr. 1, 64291 Darmstadt, Germany}
}

\author{Axel Weber}{
  address={Instituto de F\'isica y Matem\'aticas, Universidad Michoacana de San Nicol\'as de Hidalgo, \\
 Edificio C-3, Ciudad Universitaria, 58040 Morelia, Michoac\'an, Mexico}
}

\begin{abstract}
A new functional renormalization group equation for
         Hamiltonian Yang-Mills theory in Coulomb gauge is presented
         and solved for the static gluon and ghost propagators under
         the assumption of ghost dominance. The results are compared
         to those obtained in the variational approach.
\end{abstract}

\maketitle

%%%%%%%%%%%%%%%%%%%%%%%%%%%%%%%%%%%%%%%%%%%%
%% MAINMATTER
%%%%%%%%%%%%%%%%%%%%%%%%%%%%%%%%%%%%%%%%%%%%

\section{Introduction}

My talk is devoted to the application of functional renormalization group
(FRG) flows to the Hamiltonian
formulation of Yang-Mills theory in Coulomb gauge developed in our group \cite{Feuchter:2004mk}.

The advantage of the Hamiltonian formulation is its close connection
to physics. In the variational approach one makes an ansatz for the
unknown vacuum wave functional which encodes all the physics \cite{Szczepaniak:2001rg,Feuchter:2004mk}. This
ansatz can be systematically improved towards the full theory. The
price to pay is the apparent loss of renormalization group invariance.

Renormalization group invariance is naturally built-in in the
functional renormalization group approach to the Hamilton formulation
of Yang-Mills theory put forward in \cite{Leder:2010ji}. Such an
approach has the advantage of combining renormalization group
invariance with the physical Hamiltonian picture.

\section{Hamiltonian flow}

In the FRG approach the quantum theory of a field $\varphi$ is infrared regulated by adding the regulator
term \be
\label{12}
\Delta S_k [\varphi] = \frac{1}{2} \varphi \cdot R_k \cdot \varphi
\equiv \frac{1}{2} \int \frac{d^d p}{(2 \pi)^d} \varphi(p) R_k(p) \varphi(-p)
\ee
to the classical action. The regulator function
$R_k (p)$ is an effective momentum dependent mass with the
properties \be
\label{13}
\lim\limits_{p/k \to 0} R_k (p) > 0 \hk , \hk \lim\limits_{k/p \to 0} R_k
(p) = 0 \hk , \ee 
which ensures that $R_k (p)$ suppresses propagation of modes
with $p \lesssim k$ while those with $p \gtrsim k$ are unaffected and
the full theory at hand is recovered as the cut-off scale $k$ is
pushed to zero. Wetterich's flow equation for the effective action $\G_k[\phi]$ of a field $\phi$ is given by
\be
\label{18}
\partial_t \Gamma_k [\phi] = \frac{1}{2} \Tr\, 
\frac{1}{ \Gamma_k^{(2)} [\phi] + R_k } \,\dot{R}_k  ,
\ee
where
\be
\label{19}
\Gamma^{(n)}_{k, 1 \dots n} [\phi] = \frac{\delta^n 
\Gamma_k[\phi]}{\delta \phi_1 \dots \delta \phi_n}
\ee
are the one-particle irreducible $n$-point functions (proper
vertices), for reviews on gauge theories see \cite{Litim:1998nf}. The generic structure of the flow equation (\ref{18}) is
independent of the details of the underlying theory, but is a mere consequence of
the form of the regulator term (\ref{12}), i.e., that it is quadratic
in the field. By taking functional derivatives of Eq. (\ref{18}) one
obtains the flow equations for the (inverse) propagators and proper
vertices. For the two-point function this equation reads
\be\begin{split}
\label{20}
&\partial_t {\Gamma}^{(2)}_{k, 12} = \frac{1}{2} \Tr\, \dot{R}_k\, \frac{1}{\Gamma_k^{(2)} + R_k} \Biggl\{ - \Gamma^{(4)}_{k, 12} \\
&\hspace{.5cm} + \left[ \Gamma^{(3)}_{k, 1} \,\frac{1}{\Gamma_k^{(2)} + R_k} 
\,\Gamma^{(3)}_{k, 2} + \lk 1 \leftrightarrow 2 \rk \right] \Biggr\} \frac{1}{\Gamma^{(2)}_k + R_k} ,
\end{split}\ee
where all cyclic indices (summed over in the trace) have been suppressed.

In the Hamiltonian approach to Yang-Mills theory in Coulomb gauge the generating functional of static correlation functions reads 
\be
\label{23}
Z [J] = \int \D A \, \Det (- D \partial) | \psi [A] |^2 \exp (J \cdot A) \hk ,
\ee
where the integration is over transversal gauge fields $A$ and the
Coulomb gauge condition has been implemented by the usual
Faddeev-Popov method. Representing the Faddev-Popov determinant in the
standard fashion by ghost fields, $c, \cb$, 
\be
\label{264}
\Det (- D \partial) = \int \D \bar{c} \D c \, e^{- \int \bar{c} (- D \partial) c}
\ee
the underlying action reads
\be
\label{24}
S [A,\cb,c] = - \ln | \psi [A] |^2 + \int \bar{c} (- D \partial) c \hk .
\ee 
The general flow equation (\ref{20}) still holds provided that
$\phi$ is interpreted as the superfield $\phi = (A, c, \bar{c})$. The FRG flow
equations for the gluon and ghost propagators are diagrammatically
given in Figs. \ref{full flow gluon}, \ref{full flow ghost}.

\section{Approximation schemes and numerical solution}

\begin{figure}[t]
\includegraphics[width=\linewidth]{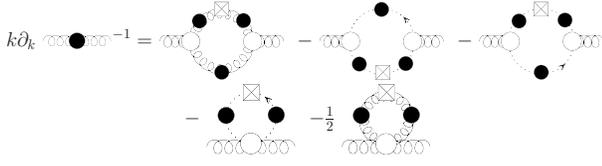}
\caption{Flow equation of the gluon propagator. The spiral and dotted lines with black circles denote the regularized gluon and ghost propagators at cutoff momentum $k$, respectively. White circles stand for proper vertices at cutoff $k$, a regulator insertion $\dot{R}_k$ is represented by a square with a cross.} 
\label{full flow gluon}
\end{figure}

\begin{figure}[t]
\includegraphics[width=\linewidth]{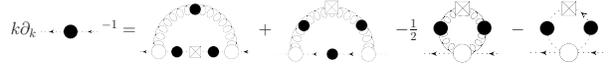}
\caption{Flow equation of the ghost propagator.} 
\label{full flow ghost}
\end{figure}

\begin{figure}[t]
\includegraphics[width=\linewidth]{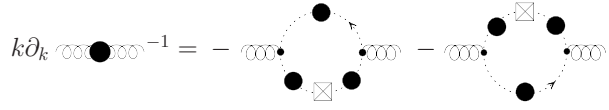}
\caption{Truncated flow equation of the gluon propagator. The bare vertices at $k=\Lambda$ are symbolized by small dots.} 
\label{trunc flow gluon}
\end{figure}

\begin{figure}[t]
\includegraphics[width=\linewidth]{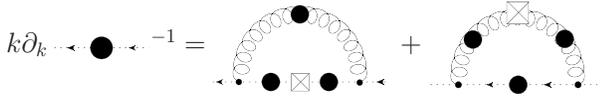}
\caption{Truncated flow equation of the ghost propagator.} 
\label{trunc flow ghost}
\end{figure}

The FRG flow equations embody an infinite tower of coupled equations
for the flow of the propagators and the proper vertices. These
equations have to be truncated to get a closed system. We shall use
the following truncation: we only keep the gluon and ghost
propagators, to wit
\be
\label{25}
\Gamma^{(2)}_{k,AA}  =  2 \omega_k (p) \,, \qquad\qquad
\Gamma^{(2)}_{k,\bar{c} c}  =  \frac{p^2}{d_k (p)} \hk ,
\ee
In addition, we keep the ghost-gluon vertex $\Gamma^{(3)}_{k,A \bar{c}
  c}$, which we approximate by the bare vertex, i.e., we do not solve its
FRG flow equation. The latter approximation is justified by Taylor's
non-renormalization theorem extended to Coulomb gauge.
The above truncation removes the tadpole diagrams
from Figs. \ref{full flow gluon}, \ref{full flow ghost}. Moreover, we shall assume infrared ghost
dominance and discard gluon loops. Then the flow equations of the
ghost and gluon propagator reduce to the ones shown in Figs. \ref{trunc flow gluon}, \ref{trunc flow ghost}.


\begin{figure}
\includegraphics[width=\linewidth]{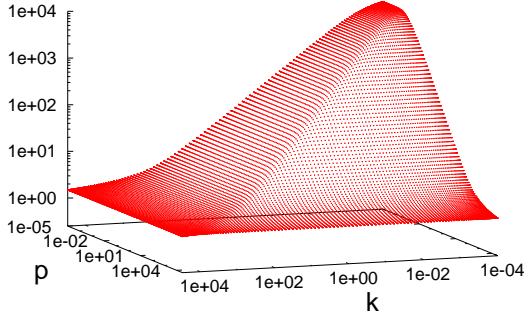}
\caption{Flow $d_k(p)$ of the ghost form factor.}
\label{ghost_flow_full.eps}
\end{figure}

\begin{figure}
\includegraphics[width=\linewidth]{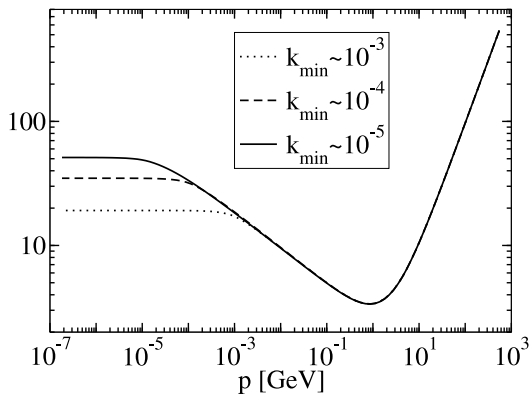}
\caption{Inverse gluon propagator $\omega$ at three minimal cutoff values $k_{min}$.\newline}  
\label{3diff_kmin_omega}
\end{figure}

\begin{figure}
\includegraphics[width=\linewidth]{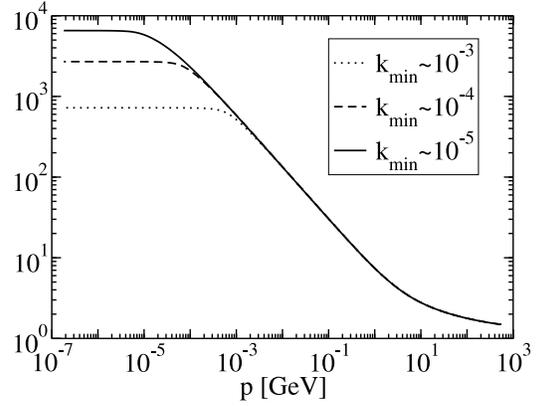} 
\caption{Inverse ghost form factor $d$ at three minimal cutoff values $k_{min}$.}  
\label{3diff_kmin_ghost}
\end{figure}

These flow equations are solved numerically using the regulators
\be\begin{split}
 & R_{A,k} (p) = 2p r_k (p) \;, \hk  R_{c,k} (p) = p^2 r_k (p) \hk , \\ & r_k (p) 
= \exp \left[ \frac{k^2}{p^2} - \frac{p^2}{k^2} \right]
\end{split}\ee
and the perturbative initial conditions at the large momentum scale $k = \Lambda$,
\be
\label{29}
d_{\Lambda} (p) = d_\Lambda = const.\;,\quad \omega_\Lambda (p) = p +
a \hk .  \ee With these initial conditions, the flow equations for the
ghost and gluon propagators are solved under the constraint of
infrared scaling for the ghost form factor. The resulting full flow of
the ghost dressing function is shown in
Fig. \ref{ghost_flow_full.eps}. As the IR cut-off momentum $k$ is
decreased, the ghost form factor $d_k(p)$ (constant at $k = \Lambda$)
builds up infrared strength and the final solution at $k = k_{min}$ is
shown in Fig. \ref{3diff_kmin_ghost} together with the one for
the gluon energy $\omega_{k_{min}}(p)$ in Fig. \ref{3diff_kmin_omega}.  It is seen that the IR
exponents, i.e., the slopes of the curves $d_{k_{min}} (p) ,
\omega_{k_{min}} (p)$ do not change as the minimal cut-off $k_{min}$
is lowered. Let us stress that we have assumed infrared scaling of the
ghost form factor but not the horizon condition $d_{k=0}^{-1}(p=0) = 0$. The latter was
obtained from the integration of the flow equation but not put in by
hand (the same is also true for the infrared analysis of the
Dyson-Schwinger equations (DSEs) following from the variational Hamiltonian
approach, i.e., assuming scaling the DSEs yield the horizon
condition).

In Coulomb gauge the inverse
ghost form factor $d^{-1}(p)$ has been shown to represent the dielectric function
of the Yang-Mills vacuum \cite{Reinhardt:2008ek}, $\epsilon (p) = d^{-
  1} (p)$. Then the so-called horizon condition $d^{- 1} (0) = 0$
implies that the Yang-Mills vacuum is a perfect dual color
superconductor.
In the variational approach one can show that the infrared exponents of the 
ghost and gluon propagators, \be
\label{7}
\omega (p\to 0) \sim 1/p^\alpha \hk , \qquad \hk d (p\to 0) \sim
1/p^\beta \hk , \ee are related by a sum rule under the assumption of
a trivial scaling of the ghost-gluon vertex
\cite{Schleifenbaum:2006bq}, 
\be
\label{8}
\alpha = 2 \beta - 1 \hk . \ee

The infrared exponents extracted from the numerical solutions of the flow equations are
\be
\label{31}
\alpha = 0.28 \, , \quad \beta = 0.64 \hk .  \ee
They satisfy the sum
rule found in \cite{Schleifenbaum:2006bq} 
but are smaller than the ones of the DSE. Moreover, the present approach allows to prove the
uniqueness of the sum rule (\ref{8}) \cite{Leder:2010ji}, analogously
to the proof in Landau gauge \cite{Fischer:2009tn}.

Replacing the propagators with running cut-off momentum scale $k$
under the loop integrals of the flow equation by the propagators of
the full theory, \be
\label{32}
d_k (p) \to d_{k = 0} (p) \hk , \hk \omega_k (p) \to \omega_{k = 0}
(p) \hk , \ee amounts to taking into account the tadpole diagrams
\cite{Leder:2010ji}. Then the flow equations can be analytically
integrated and turn precisely into the DSEs obtained in the
variational approach to the Hamiltonian formulation of Yang-Mills
theory \cite{Feuchter:2004mk}, with explicit UV regularization by
subtraction. This establishes the connection between these two
approaches and highlights the inclusion of a consistent UV
renormalization procedure in the present approach.

The above results encourage further studies, which include the
flow of the potential between static color sources as well as dynamic
quarks.

%%%%%%%%%%%%%%%%%%%%%%%%%%%%%%%%%%%%%%%%%%%%%%%%
%% BACKMATTER
%%%%%%%%%%%%%%%%%%%%%%%%%%%%%%%%%%%%%%%%%%%%%%%%

%\begin{theacknowledgments}
%\end{theacknowledgments}




%\bibliography{sample}


\end{document}